\documentclass[12pt]{iopart}

\usepackage{iopams}
\usepackage{graphicx}
\usepackage{setspace}
\usepackage{bm}
\usepackage[dvips,usenames]{color}
\usepackage{fancyhdr}
\scrollmode
\eqnobysec

\begin{document}

\fancyhf{}
\fancyhead[L]{\textit{\nouppercase{}Thermoelectric efficiency}}
\fancyhead[R]{\nouppercase{G Benenti} \textit{\nouppercase{et al}}}
\fancyfoot[C]{\thepage}

\title[Thermoelectric efficiency]{Thermoelectric efficiency in momentum-conserving systems}

\author{Giuliano Benenti$^{1,2}$, Giulio Casati$^{1,2}$ and
Carlos Mej\'{\i}a-Monasterio$^{3,4}$}

\address{$^1$ CNISM and Center for Nonlinear and Complex Systems,
  Universit\`a dell'Insubria, via Vallegio 11, 22100 Como, Italy}
\address{$^2$ Istituto Nazionale de Fisica Nucleare, Sezione di
  Milano, via Celora 16, 20133 Milano, Italy}
\address{$^3$ Laboratory of Physical Properties TAGRALIA,
Technical University of Madrid, Av. Complutense s/n, 28040 Madrid, Spain}
\address{$^4$Department  of Mathematics and  Statistics, University  of 
Helsinki,  P.O.   Box  68  FIN-00014  Helsinki,  Finland}

\eads{\mailto{giuliano.benenti@uninsubria.it}, \mailto{giulio.casati@uninsubria.it}, \mailto{carlos.mejia@upm.es}}

\begin{abstract}
We show that for a two-dimensional gas of elastically interacting particles
the thermoelectric efficiency reaches the Carnot efficiency in the thermodynamic limit.
Numerical simulations, by means of the multi-particle collision dynamics 
method, 
show that this result is robust under perturbations.
That is, the thermoelectric figure of merit remains large 
when momentum conservation is broken by weak noise. 
\end{abstract}

\vspace{2pc}

\submitto{NJP}
%\maketitle

%\tableofcontents

\section{Introduction}
\label{sec:intro}

Understanding  and  controlling  the behaviour  of  out-of-equilibrium
systems  is  one  of   the  major  challenges  of  modern  statistical
mechanics.   From a  fundamental point  of view,  the challenge  is to
understand the origin  of macroscopic transport phenomenological laws,
such as diffusion equations, in terms of the properties of microscopic
dynamics,  typically  nonlinear  and  chaotic~\cite{lepri,dhar}.   The
problem is extremely complex for  coupled flows, so far barely studied
from                           the                          viewpoint
of statistical mechanics and dynamical systems
\cite{vollmer,carlos2001,carlos2003,maes,carlos2008,casati2009,wang2009,saito2010,BC11,lepri2012}.
In  particular,  it  is   of  primary  importance  for  thermoelectric
transport~\cite{dresselhaus,snyder,shakuori,dubi}  to  gain a  deeper
understanding  of the microscopic  mechanisms leading  to a
large thermoelectric efficiency; see Ref.~\cite{RMP} for a review
on the fundamental aspects of heat to work conversion.

Within    linear   response,    and   for    time-reversal   symmetric
systems~\footnote{Thermodynamic bounds on  efficiency for systems with
  broken       time-reversal        symmetry       are       discussed
  in~\cite{benenti2011,saito2013,vinitha2013,seifert2013}}  both  the  maximum  thermoelectric
efficiency    and    the   efficiency    at    the   maximum    output
power~\cite{vandenbroeck,esposito2009,schulman,esposito2010,seifert}
are  monotonous growing  functions of  the so-called  figure  of merit
$ZT=(\sigma S^2/\kappa)T$, which is a dimensionless combination of the
main  transport  coefficients  of  a  material, that  is,  the  electric
conductivity  $\sigma$,  the  thermal  conductivity $\kappa$  and  the
thermopower (Seebeck coefficient) $S$, and of the absolute temperature
$T$.    The  maximum  efficiency   reads  $\eta_{\rm   max}=  \eta_C\,
\frac{\sqrt{ZT+1}-1}{\sqrt{ZT+1}+1}$,  where  $\eta_C$  is the  Carnot
efficiency, while the efficiency at maximum output power $P_{\rm max}$
is  given  by  $\eta(P_{\rm  max})=\frac{\eta_{C}}{2}\frac{ZT}{ZT+2}$.
Thermodynamics only imposes $ZT\ge  0$ and $\eta_{\rm max}\to \eta_C$,
$\eta(P_{\rm max})\to \frac{\eta_C}{2}$ when $ZT\to\infty$.

Since the  different transport coefficients are  interdependent, it is
very  difficult to  find  microscopic mechanisms  which could  provide
insights   to   design  materials   with   large   $ZT$.   While   for
non-interacting   models   it   is   well   understood   that   energy
filtering~\cite{mahansofo,linke1,linke2} allows us to reach the Carnot
efficiency,   very   little   is   known   for   interacting   systems
\cite{carlosAIP}.  It has  been recently shown~\cite{benenti2013} that
the thermoelectric figure of  merit $ZT$ diverges in the thermodynamic
limit  for  systems  with  a  single relevant  conserved  quantity,  an
important  example being that of momentum-conserving systems,
with total momentum being the only relevant constant of motion.  While
the   mechanism    is   generic,   it   has    been   illustrated   in
Ref.~\cite{benenti2013} only  for a toy model, i.e.,  a diatomic chain
of hard-point elastically colliding particles.

In this paper,  we show by means of  extensive multi-particle collision
dynamics simulations that the momentum-conservation mechanism leads to
the  Carnot efficiency  in the  thermodynamic limit  also in  the more
realistic    case    of    two-dimensional    elastically    colliding
particles. Furthermore, we show that this mechanism leads to a 
significant enhancement of the thermoelectric figure of merit even when
the  momentum conservation is  not exact  due to  the existence  of an
external noise. This robustness is particularly relevant in experiments
for which  inelastic or incoherent processes are unavoidable to some
extent.  In this case, the figure  of merit saturates with the size of
the system  to a value  higher, the weaker  is the noise.  Finally, we
discuss the validity range of linear response.
%show  that the  validity range  of  linear response  shrinks with  the
%system  size, so  that, as  expected  on general  grounds, the  Carnot
%efficiency is obtained in the  limit of zero current, corresponding to
%reversible transport (zero entropy production) and zero output power.

The  paper  is  organised  as follows.   In  Sec.~\ref{sec:theory},
in order to  make the  paper self-contained, we review the theoretical
argument of  Ref.~\cite{benenti2013} explaining the  divergence of the
thermoelectric  figure of merit  $ZT$ in  the thermodynamic  limit for
systems   with   a   single   relevant   constant   of   motion.    In
Sec.~\ref{sec:method} we explain  our out-of equilibrium multi-particle
collision dynamics  simulations.  Our numerical  results are presented
in  Sec.~\ref{sec:discussion}.   We finish  with  concluding remarks  in
Sec.~\ref{sec:conclusions}.

\section{Theoretical argument} 
\label{sec:theory}

\subsection{Linear response irreversible thermodynamics}

The equations connecting fluxes and thermodynamic forces within linear
irreversible thermodynamics read as
follows~\cite{callen,degrootmazur}:
\begin{equation} \label{phenomeqs}
\left(
\begin{array}{c}
J_\rho\\
J_u
\end{array}
\right) = \left(
\begin{array}{cc}
L_{\rho \rho} & L_{\rho u} \\
L_{u \rho} & L_{u u}
\end{array}
\right) \left(
\begin{array}{c}
-\nabla(\beta\mu)\\
\nabla \beta
\end{array}
\right) \ ,
\label{eq:lresponse}
\end{equation}
where $J_\rho$ and $J_u$ are the particle and energy currents, $\mu$
the chemical potential and $\beta=1/T$ the inverse temperature (we set
the Boltzmann constant $k_B=1$).  The kinetic coefficients $L_{ij}$ (
$i,j=\{\rho,u\}$), are related to the familiar transport coefficients
as
\begin{equation} \label{transport}
\sigma=\frac{L_{\rho\rho}}{T},
\quad\kappa=\frac{1}{T^2}\frac{\det {\bm L}}{L_{\rho\rho}},
\quad S=\frac{1}{T}\left(\frac{L_{\rho u}}{L_{\rho\rho}}-\mu\right),
\end{equation}
where ${\bm L}$ denotes the (Onsager) matrix of kinetic coefficients
and we have set the electric charge of each particle $e=1$.
Thermodynamics imposes $\det {\bm L}\ge 0$, $L_{\rho\rho}\ge 0$,
$L_{uu}\ge 0$; $L_{u\rho}=L_{\rho u}$ follows from the Onsager
reciprocity relations.  The thermoelectric figure of merit reads
\begin{equation} \label{ZT}
ZT=\frac{(L_{u\rho}-\mu L_{\rho\rho})^2}{\det {\bm L}}
=\frac{\sigma S^2}{\kappa}\,T.
\end{equation}

Furthermore, the Green-Kubo formula expresses the kinetic coefficients
in  terms  of  correlation  functions  of  the  corresponding  current
operators, calculated at thermodynamic equilibrium~\cite{kubo,mahan}:
\begin{equation}
L_{ij} = \lim_{\omega\to 0} {\rm Re} L_{ij} (\omega),
\end{equation}
where
\begin{equation}
%\begin{array}{c}
L_{ij}(\omega)\equiv \lim_{\epsilon\to 0}
\int_0^\infty dt e^{-i(\omega-i\epsilon)t}
%\\
%\\
%\times
\lim_{\Omega \to\infty}
\frac{1}{\Omega}
\int_0^\beta d\tau\langle {J}_i {J}_j (t+i\tau)\rangle,
%\end{array}
\label{eq:kubo}
\end{equation}
where $\langle  \; \cdot \;\rangle  = \left\{{\rm tr}\left[(\;\cdot\;)
    \exp^{-\beta    H}\right]\right\}/{\rm    tr}    \left[\exp(-\beta
  H)\right] $ denotes the equilibrium expectation value at temperature
$T$  and $\Omega$  is the  system's volume.   Within the  framework of
Kubo's linear  response approach, the real part  of $L_{ij}(\omega)$ can
be decomposed into a  singular contribution at zero frequency  and  a  regular  part
$L_{ij}^{\rm reg}(\omega)$ as
\begin{equation}
{\rm Re} L_{ij}(\omega)=
2\pi {\cal D}_{ij}\delta(\omega)+L_{ij}^{\rm reg}(\omega) \ .
\label{eq:reLij}
\end{equation}
The coefficient of the singular part defines the generalized Drude
weights ${\cal D}_{ij}$~\footnote{For $i=j=\rho$, we have the 
conventional Drude weight
${\cal    D}_{\rho\rho}$.}, which can be expressed as~\footnote{See 
Ref.~\cite{prosen2013}    for   a   detailed
  discussion  and derivation  of  Eq.~(\ref{eq:drudeinfinite}).}
\begin{equation}
{\cal D}_{ij}=\lim_{t\to\infty}\lim_{l\to \infty}
\frac{1}{2\Omega(l) t}
\int_0^{t} dt' \langle J_i(t') J_j(0) \rangle \ ,
\label{eq:drudeinfinite}
\end{equation}
where in the volume $\Omega(l)$  we have explicitly written the 
dependence on the system size $l$ along the direction of the 
thermodynamic flows. 
%The  important remark 
%from  Eq.~\eref{eq:drudeinfinite}  is  that  
Non-zero  Drude  weights,
${\cal    D}_{ij}\ne    0$,    
are    a    signature    of    ballistic
transport~\cite{zotos,zotosreview,garst,heidrich-meisner},  namely  in
the  thermodynamic  limit  the  kinetic coefficients  $L_{ij}$  scale
linearly with  the system size $l$.
As a consequence, the thermopower $S$ does not scale with $l$.

\subsection{Conservation laws}
\label{conslaw}

We now discuss the influence  of conserved quantities on the figure of
merit  $ZT$. Making use  of Suzuki's  formula~\cite{suzuki}  for the
currents    $J_\rho$   and    $J_u$,    one can   generalize    Mazur's
inequality~\cite{mazur} by  stating that, for a system  of finite size
$l$ (along the direction of the flows),
\begin{equation}
C_{ij}(l)\equiv \lim_{t\to\infty} C_{ij}(t) =
\lim_{t\to\infty} \frac{1}{t}
\int_0^{t} dt' \langle J_i(t') J_j(0) \rangle
=
\sum_{n=1}^M
\frac{\langle J_i Q_n \rangle
\langle J_j Q_n \rangle}{\langle Q_n^2\rangle},
%\quad (i,j=\rho,u),
\label{eq:suzuki}
\end{equation}
where for readability, in the right hand side of the equation we have
omitted the dependence on $l$.  The summation in Eq.~\eref{eq:suzuki}
extends over all the $M$ constants of motion $Q_n$, which are
orthogonal, $\langle Q_n Q_m \rangle = \langle Q_n^2 \rangle
\delta_{n,m}$, and relevant for the considered flows. That is,
$\langle J_\rho Q_n \rangle\ne 0$ and $\langle J_u Q_n \rangle\ne 0$.

From Eq.~(\ref{eq:suzuki}) one can define the finite-size generalized Drude
weights as
\begin{equation} \label{drude-1}
D_{ij}(l)\equiv \frac{1}{2\Omega(l)}\,C_{ij}(l) \ .
\end{equation}
Therefore, the presence of relevant conservation laws directly implies
that the finite-size generalized Drude weights are  different   from  zero.
If the thermodynamic   limit
$l\to\infty$ can be taken after the long-time limit $t\to\infty$,
so that the generalized Drude coefficients can be written as 
\begin{equation} \label{drude-2}
{\cal D}_{ij}=\lim_{l\to\infty} D_{ij}(l), \ 
\label{eq:suzukilimit}
\end{equation}
and moreover ${\cal D}_{ij}\ne 0$, 
then we can  conclude that the presence of  relevant conservation laws
yield non-zero  generalized Drude weights, which in turn imply that
transport is ballistic.
%The exchangeability  of order of the limits 
%the basis  of Suzuki's approach to thermodynamics.   
%When possible, it
%means  that the  dynamics  of the  small  system is  the  same as  the
%dynamics of the big system. However,  this is not always the case and,
%as a matter of fact, taking the limit $l\to\infty$ before $t\to\infty$
%is a fundamental principle of statistical mechanics. 
We point out that, in contrast to Eq.~(\ref{eq:suzukilimit}), 
one should take the thermodynamic limit $l\to\infty$ before 
the long-time limit $t\to\infty$. While it remains an interesting open problem
for which classes of models the two limits commute~\footnote{See
  Ref.~\cite{prosen2013}  for a proof  of the  commutation of  the two
  limits for a class of quantum spin chains.},
numerical evidence suggests that it is possible to commute the limits 
for the models considered in Ref.~\cite{benenti2013}
and in the present paper. 
%Therefore, our theoretical argument below only applies to systems for which the 
%two limits commute.

%Assuming that the two limits $l\to\infty$ and $t\to 0$ commute.
Let us first consider the case in which there
is a single relevant constant of motion, $M=1$.
We can see from Suzuki's formula, Eq.~\eref{eq:suzuki}, that
the ballistic contribution to
$\det {\bm L}$ vanishes, since it is proportional to ${\cal
  D}_{\rho\rho}{\cal D}_{uu}-{\cal D}_{\rho u}^2$, which is
zero from \eref{eq:suzuki} and \eref{eq:suzukilimit}.  Hence, $\det
{\bm L}$ grows only due to the contributions involving the regular
part in Eq.  (\ref{eq:reLij}), i.e., slower than $l^2$, which in turn
imply that the thermal conductivity $\kappa\sim \det{\bm
  L}/L_{\rho\rho}$ grows sub-ballistically.  Furthermore, since $\sigma\sim L_{\rho\rho}$ is
ballistic and $S\sim l^0$, we can
conclude that
\begin{equation}
ZT=\frac{\sigma S^2 T}{\kappa} \propto \frac{l}{k}\ .
%\lesssim l \ .
\end{equation}
Thus $ZT$ diverges in the thermodynamic limit $l\to\infty$.

The situation  is drastically different if  $M>1$, as it  would be the
case  for integrable  systems, where typically the number of 
orthogonal relevant constants of motion equals the 
number of degrees of freedom. 
In that case, due to the Schwartz inequality,
\begin{equation}
D_{\rho\rho}D_{uu}-D_{\rho u}^{2}=
||{\bm x}_\rho||^2 ||{\bm x}_u||^2-
\langle {\bm x}_\rho , {\bm x}_u \rangle^2 \ge 0,
\end{equation}
where
\begin{equation}
{\bm x}_i= (x_{i1},...,x_{iM})=\frac{1}{\sqrt{2\Omega(l)}}
\left(\frac{\langle J_i Q_1\rangle}{\sqrt{\langle Q_1^2 \rangle}},...,
\frac{\langle J_i Q_M\rangle}{\sqrt{\langle Q_M^2 \rangle}}
\right),
\end{equation}
and $\langle {\bm  x}_\rho , {\bm x}_u \rangle  = \sum_{k=1}^M x_{\rho
  k} x_{uk}$.  The  equality arises only in the  exceptional case when
the vectors ${\bm x}_{\rho}$ and ${\bm x}_u$ are parallel.  Hence, for
$M>1$ we expect,  in general, $\det {\bm L}\propto  l^2$, so that heat
transport is ballistic and $ZT\sim l^0$.

\section{Momentum-conserving gas of interacting particles}
\label{sec:method}

In this section we analyse  the consequences of our analytical results
in a  two-dimensional gas of  interacting particles.
We consider a  gas of point-wise particles in a
rectangular two-dimensional box of length $l$ and width $w$. The gas
container is placed in contact with two particle reservoirs at $x=0$ and
$x=l$, through openings of the same size as the width $w$ of the box.
In the transversal direction the particles are subject to periodic
boundary conditions.

The dynamics of the particles in  the system are solved by the method
of Multi-particle Collision Dynamics (MPC) \cite{kapral}, introduced as a 
stochastic model  to study solvent dynamics. The MPC simplifies the
numerical simulation  of interacting particles by  coarse graining the
time and space  at  which  interactions occurs. MPC correctly captures
the hydrodynamic equations \cite{padding,gompper}.  It   has  been
successfully applied to model steady shear flow situations in colloids
\cite{hecht},   polymers   \cite{nikoubashman},   vesicles  in   shear
flow~\cite{noguchi}, colloidal  rods \cite{ripoll}, and  more recently
to  study the  steady-state of  a gas  of particles  in  a temperature
gradient \cite{ripoll12}.

Under MPC dynamics the system  evolves in discrete time steps, consisting
on  free  propagation during  a  time  $\tau$,  followed by  collision
events.   During propagation,  the coordinates  $\vec{r}_{i}$  of each
particle  are  updated   as
\begin{equation}
\vec{r}_{i}  \rightarrow  \vec{r}_{i}  +
\vec{v}_{i}\tau \ ,
\end{equation}
where $\vec{v}_{i}$ is the particle's velocity. For the collisions the
system's  volume  is  partitioned  in  identical cells  of  linear  size
$a$. Then, the velocities of the $\mathcal{N}$ particles found 
in the same cell are
rotated  with respect  to  the center  of  mass velocity  by a  random
angle.  In  two  dimensions, 
rotations by an angle $+\alpha$ or $-\alpha$ with equal probability 
$p(+\alpha)=p(-\alpha)=1/2$ 
are performed.
%one performs choosing  a  fixed  angle  $\alpha$  and
%performing clockwise or anti-clockwise  rotations randomly with equal weights, 
%is enough to guarantee global time reversibility. 
The velocity updating after
a collision event reads
\begin{equation}
\vec{v}_{i} \rightarrow \vec{V}_{\mathrm{CM}} +
\hat{\mathcal{R}}^{\pm\alpha}\left(\vec{v}_{i} -
  \vec{V}_{\mathrm{CM}}\right) \ ,
\end{equation}
where                     $\vec{V}_{\mathrm{CM}}                     =
\frac{1}{\mathcal{N}}\sum_{i=1}^\mathcal{N} \vec{v}_{i}$ is the center
of   mass    velocity   and   $\hat{\mathcal{R}}^{\theta}$    is   the
two-dimensional rotation operator  of angle $\theta$.  Furthermore, to
guarantee Galilean invariance, the  collision grid is shifted randomly
before  each  collision  step.   It  has been  shown  that  for  these
dynamics, the equation of state of the gas of particles corresponds to
that  of an  ideal  gas \cite{kapral}.   Moreover,  the time  interval
between successive collisions $\tau$  and the collision angle $\alpha$
tune the strength of the  interactions and consequently
affects the transport coefficients  of the gas of
particles. When $\alpha$ is a multiple of $2\pi$, the particles do not
interact, propagating ballistically from one reservoir to the other as
they cross the system. For  any other value of $\alpha$, the particles
interact,  exchanging momentum during  the collision  events. 
The value $\alpha=\pi/2$ corresponds to  the  most efficient  
mixing of the particle momenta.
Note  that by construction,  the collision preserve  the total
energy and total momentum of the gas of particles.

From the reservoir $k$ ($k=L,R$ for the left and the right reservoir),
particles of mass $m$ enter  the system at rate $\gamma_k$ obtained by
integration of the appropriate canonical distribution to give
\begin{equation} \label{eq:gamma}
\gamma_k \ = \ \frac{w}{\left(2\pi m\right)^{1/2}}\rho_k 
T_k^{1/2} \ ,
\end{equation}
where $\rho_k$ and $T_k$ are 
the particle density temperature. 
Assuming that the particles in the reservoirs behave as ideal gas, the
particle  injection rate  is  related  to the  value  of the  chemical
potential $\mu_k$ of the reservoir $k$ as
\begin{equation} \label{eq:chempot}
\mu_k = T_k \ln\left(\frac{\gamma_k}{T_k^{3/2}}\right) + 
\mu_0 \ , 
\end{equation}
with $\mu_0$ an arbitrary constant whose value does not qualitatively 
modify the results discussed in this paper; hereafter we set 
$\mu_0$ in such a way that $\mu=0$~\footnote{This arbitrariness is intrinsic in classical mechanics
and can only be removed by means of semiclassical arguments, 
see Ref.~\cite{saito2010}.}. 
Whenever a particle from the system crosses the boundary which separates
the system from reservoir $k$, it is removed (absorbed in the 
reservoir), i.e., it has no further effects on the evolution of the system.

\section{Discussion of numerical results}
\label{sec:discussion}

We have  numerically studied the nonequilibrium transport  of the model
defined in Sec.~\ref{sec:method}, coupled to two ideal
particle reservoirs.   The nonequilibrium state is  imposed by setting
the values  of $T$ and $\mu/T$ in  the reservoirs to different
values, meaning that  from each of the reservoirs,  the particles 
are injected into  the system at different rates  and with a different
distribution  of their  velocities.   Out of  equilibrium the  kinetic
coefficients  $L_{ij}$ can be  computed, in the linear response regime, 
by  direct measurement  of the
particle and energy currents in the system.  Using (\ref{phenomeqs}), it
is  enough to  perform two  nonequilibrium numerical  simulations: one
with $T_L\ne T_R$ and $\mu_L/T_L=\mu_R/T_R$,  and one
with $T_L= T_R$ and $\mu_L/T_L\ne\mu_R/T_R$.  
%We parametrize the gradients in terms of the difference $\Delta$, 
%so that
%to obtain a temperature gradient  of $\Delta/l$, the values of
In the first simulation the reservoirs' temperatures are set to  
$T_L = T - \Delta T/2$ and $T_R
= T + \Delta T/2$, so that the temperature gradient is given by
$\Delta T/l$, while  $\mu_L/T_L = \mu_R/T_R$. Conversely, 
in the second simulation we set  $T_L  = T_R=T$  and  using
\eref{eq:chempot}, we  set the particle injection  rates $\gamma_L$ and
$\gamma_R$ so that $\Delta\left({\mu}/{T}\right)=
\mu_L/T_L - \mu_R/T_R =(\mu_L-\mu_R)/T$.

In all simulations  the mean particle density and  mean temperature in
the reservoirs  was set to $n=N/lw=22.75$  ($N$ is the  mean number of
particles)  and  $T=1$,  respectively.   
We parametrize the gradients in terms of a single parameter by setting,
$\Delta T=\Delta \left({\mu}/{T}\right)\equiv \Delta$ 
(in units where $k_B=e=1$).
The rotation  angle  for  the
collisions  in the  MPC scheme  was set  to $\alpha  =  \pi/2$, unless
otherwise specified.   The length  of the collision  cells in  the MPC
scheme was set to $a=0.1$ and the time step to $\tau=0.25$.  For these
values  and  small  $\Delta$  the system  exhibits  reasonably  linear
temperature  and chemical potential  profiles in  the bulk,  with some
nonlinear boundary layer near the contacts, arising from the fact that
the mean free  path of the particles near  the boundaries is different
than in  the bulk \footnote{The  MPC collisions at the  boundaries are
  implemented  without  taking  into  account  the  particles  in  the
  reservoirs},  yielding  a  contact  resistance~\cite{ripoll12}.   We
performed numerical simulations with  up to $\Omega=10^3$ ($l=500$ and
$w=2$),  so  that  systems  with   mean  number  of  particles  up  to
$N=4.55\times 10^4$ were considered.

\subsection{Linear response transport}

Using  the   Suzuki's  formula  \eref{eq:suzuki},   the  current-current
correlation functions  $C_{ij}(t)$ can be  obtained analytically.  The
particle  current is  $J_\rho=\sum_{i=1}^{N} v_{x,i}$  and  the energy
current    $J_u=\frac{1}{2}\,   m\sum_{i=1}^{N}    \left(v_{x,i}^2   +
  v_{y,i}^2\right)  v_{x,i}$ where the  coordinate $x$  corresponds to
the direction of the  thermodynamic gradients, thus the direction of
the flows.

Furthermore, for the MPC model  there exists a single relevant constant
of motion, namely the  $x$-component of the
total momentum $Q_1=p_x=m \sum_{i=1}^N v_{x,i}$.
The other constants of motion,
i.e. momentum in the transverse direction, energy and number of particles,
are irrelevant since they are orthogonal to the considered flows.
Therefore, in this case $M=1$.

Applying  Eq.~(\ref{eq:suzuki})  and  integrating over  the  equilibrium
state at temperature $T$ and fixed number of particles $N$, we obtain
that the finite-size correlators are
\begin{equation} \label{correlators}
C_{\rho \rho}(l)  =  \frac{NT}{m} \ , \quad
C_{\rho u}(l)  =  \frac{2NT^2}{m} \ , \ \mathrm{and} \quad 
C_{u u}(l)  =  \frac{4NT^3}{m} \ .
\end{equation}
 
\begin{figure}[!t]
  \centerline{\includegraphics*[width=0.7\textwidth]{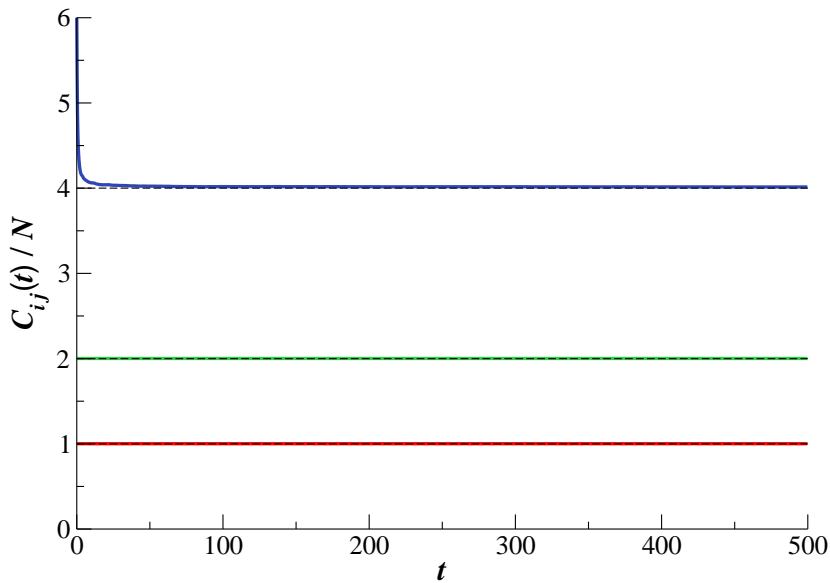}}
  \caption{Equilibrium     current-current    correlation    functions.
From bottom to top:
    $C_{\rho\rho}(t)$ (in  red), $C_{\rho u}(t)$ (in  green) and $C_{u
      u}(t)$ (in  blue), averaged  over the ensemble  of realisations.
    The   dashed  horizontal   lines   indicate  their   corresponding
    analytical values from Eq.~\eref{correlators}.}
\label{fig1}
\end{figure}

To  verify Eq.~\eref{eq:suzuki} we  have numerically  computed the
equilibrium current-current time correlation functions for the isolated
system, averaged  over an  equilibrium ensemble of  initial conditions
with $N=1000$ particles of mass  $m=1$ and temperature $T=1$. A square
container  of size  $l=2$  and periodic  boundary  conditions in  both
directions  was  considered. The  results,  shown in  Fig.~\ref{fig1},
verify  our  the  analytical  expressions. Note that 
the initial values $C_{\rho\rho}(0)$ and $C_{\rho u}(0)$ of the 
time-averaged correlation  functions
$C_{\rho \rho}(t)$  and $C_{\rho  u}(t)$ are equal to their asymptotic 
values $C_{\rho \rho}(l)=\lim_{t\to\infty} C_{\rho\rho}(t)$ and
$C_{\rho u}(l)=\lim_{t\to\infty} C_{\rho u}(t)$.
On the other hand, it is easy to compute analytically $C_{uu}(0)=6NT^3/m $, 
and numerical data show that $C_{uu}(t)$ converges 
algebraically to its asymptotic value $C_{uu}(l)=\lim_{t\to\infty} C_{u u}(t)=
4NT^3/m $. This asymptotic behaviour may be due to the slow decay of
the energy hydrodynamic modes. 

%\subsubsection{Singular limit of $\det L$.}

\begin{figure}[!t]
\begin{center}
  \centerline{\includegraphics*[width=0.8\textwidth]{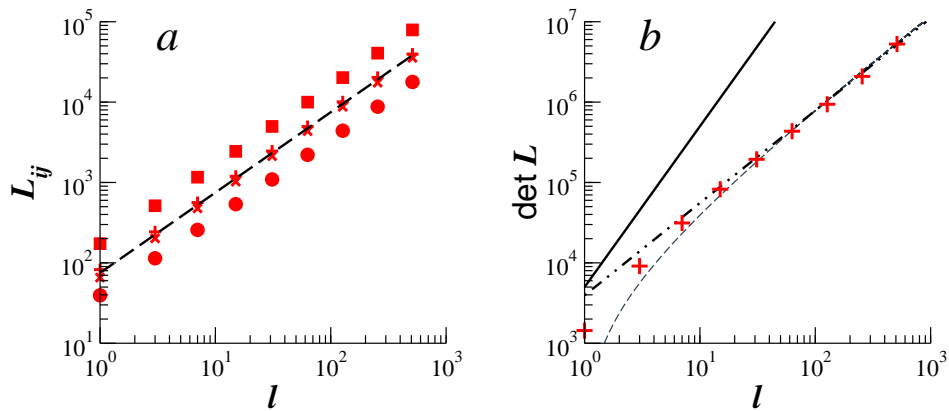}}
\caption{Dependence of the kinetic coefficients on the length of the
  system $l$, obtained from nonequilibrium simulations with  $\Delta=0.00625$.
In panel {\it a} we show the kinetic coefficients $L_{\rho \rho}$
(circles), $L_{u \rho}$ (crosses), $L_{\rho u}$ (pluses) and $L_{u
  u}$ (squares). The dashed line
stands for the linear scaling $\sim l$. In panel {\it b} we plot the
determinant of the Onsager matrix $L$ (symbols), as a function of the
length of the system. The different curves correspond to
the scalings $\sim l^2$ (solid), $\sim
l\log(l)$ (dashed) and  $\sim l^{1.15}$ (dotted-dashed).
Parameter values: $m=1$, $T=1$, $n=22.75$, $\alpha=\pi/2$,
$w=2$, $a=0.1$, $\tau=0.25$.
\label{fig2}}
\end{center}
\end{figure}

Equation  \eref{correlators} also  shows  that the  dependence of  the
correlations on the  size $l$ comes exclusively through  the number of
particles $N$. The thermodynamic limit requires keeping the density of
particles fixed, so that the number of particles has to scale linearly
with  the  volume  of  the system:  $N\propto  \Omega(l)=lw$.   Therefore,
Eqs.~\eref{correlators} imply  that the finite-size  generalized Drude
weights of  Eq.~\eref{drude-1} do not  scale with $l$.
Using Eqs.~\eref{eq:suzuki}-\eref{drude-2} we  obtain, for
fixed $w$, the generalized Drude weights
\begin{equation} \label{drude-3}
{\cal D}_{\rho \rho} = \frac{nT}{2m} \ , \quad  {\cal D}_{\rho u} =
{\cal D}_{u \rho} = \frac{nT^2}{m} \ , \ \mathrm{and} \quad  
{\cal D}_{u u} = \frac{2nT^3}{m} \ .
\end{equation}
As a consequence of the finiteness of the Drude weights, the transport
is  ballistic, meaning  that all  kinetic coefficients  $L_{ij}$ scale
linearly with the size of  the system: $L_{ij}\sim l$. This prediction
is  confirmed  by  the  numerical   results  shown  in  panel  $a$  of
Fig.~\ref{fig2}.

More  importantly,  as discussed  in  Sec.~\ref{conslaw},  due to  the
conservation  of total  momentum,  the ballistic  contribution to  the
determinant of the  Onsager matrix is zero. Indeed,  it can be readily
seen  from Eq.~\eref{drude-3}  that ${\cal  D}_{\rho  \rho}{\cal D}_{u
  u}-{\cal D}_{\rho  u}^2=0$.  Hence a scaling  $\det({\bm L})$ slower
than $l^2$ is expected.  From the nonequilibrium numerical simulations
the scaling of the determinant  with $l$ is consistent with $\det({\bm
  L})\approx  l^{1.15}$ (dotted-dashed curve  in Fig.~\ref{fig2}-$b$).
It is  worthwhile recalling that different analytical  methods such as
mode   coupling  theory  and   hydrodynamics  predict,   for  momentum
conserving systems  in two  dimensions, a logarithmic  divergence 
of the thermal conductivity with
the size of the system \cite{lepri,dhar}. Therefore, one should expect
that $\det({\bm L}) \sim l\log(l)$. We show in Fig.~\ref{fig2}-$b$
(dashed curve), that such scaling is also consistent with our
numerical results, though deviations are larger than for the algebraic
behaviour at small system sizes. Since we have no reason to expect an
algebraic sub-ballistic behaviour of the heat conductivity, we will
assume in what follows that its behaviour is logarithmic.

\subsection{Strong enhancement of $ZT$}

From Eqs.~\eref{transport} and  Eqs.~\eref{drude-3} we obtain that the
electric conductivity also scales linearly with the size of the system:
\begin{equation} \label{sigma}
\sigma = \frac{A n}{m}\,l \ ,
\end{equation}
with $A$ constant.
The dependence on $l$ of the Seebeck coefficient cancels out to give,
asymptotically in $l$,
\begin{equation}   \label{S}  S=\frac{1}{T}\left(\frac{{\cal  D}_{\rho
        u}}{{\cal D}_{\rho\rho}}   -\mu\right)=2 \ .
\end{equation}
Since the ballistic contribution to $\det({\bm L})$ vanishes,
i.e. ${\cal D}_{\rho\rho}{\cal D}_{uu}
-{\cal D}_{\rho u}^2=0$, 
we cannot derive an explicit expression for the heat conductivity 
$\kappa$. However, as discussed in the previous section, for
momentum conserving two-dimensional systems it is predicted that $\kappa$ 
diverges logarithmically with respect to the size of the system: $\kappa\sim \log(l)$.

\begin{figure}[!t]
\begin{center}
  \centerline{\includegraphics*[width=1\textwidth]{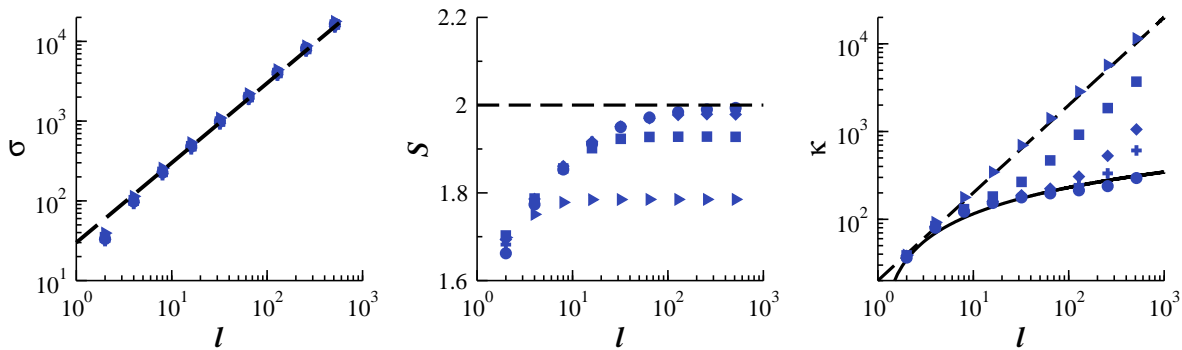}}
\caption{Transport coefficients as a function o  the
length of  the system $l$, for different thermodynamic gradients,
$\Delta=0.00625$ (circles), $0.0125$      (pluses), $0.025$
(diamonds), $0.1$  (squares) and $0.4$  (triangles) and the 
values  of other parameters as in the caption of Fig.~\ref{fig2}. In panel
{\it a} the dashed line corresponds to Eq.~\eref{sigma} with 
$A=\pi/4$ and in
panel {\it b} to $S=2$. In panel {\it c} the dashed line stands for
linear scaling $\sim
l$, while the solid line corresponds to $\log(l)$.
\label{fig3}}
\end{center}
\end{figure}

Fig.~\ref{fig3} shows the dependence  of the transport coefficients on
the size  of the system,  for different values of  the thermodynamic
forces.    The   electric   conductivity   verifies   Eq.~\eref{sigma}
independently  of  the  value   of  the  thermodynamic  force  $\Delta$,
with the constant $A=\pi/4$.
Instead, the Seebeck coefficient shows a clear dependence on $\Delta$,
verifying Eq.~\eref{S} (asymptotically in  $l$) only in the limit of
small  forces (in  Fig.~\ref{fig3}, $S$  is shown  for  $\mu=0$).
We have found that $S$ converges to the value $S=2$ 
predicted by \eref{S} as $1/\Delta$.

The heat  conductivity $\kappa$ exhibits the  logarithmic behaviour up
to  a size  $l=l^\star$  dependent  on the  strength  $\Delta$ of  the
thermodynamic   forces.   For   any  value   of  $\Delta$,   the  heat
conductivity grows  as $\kappa\sim \log(l)$,  for $l >  l^\star$.  The
smaller  the  $\Delta$  the  larger  the  range  of  validity  of  the
logarithmic  $\kappa$  is.  We  have  obtained  numerically  that  the
characteristic length $l^\star$ grows linearly with $1/\Delta$.

\begin{figure}[!t]
\begin{center}
  \centerline{\includegraphics*[width=0.8\textwidth]{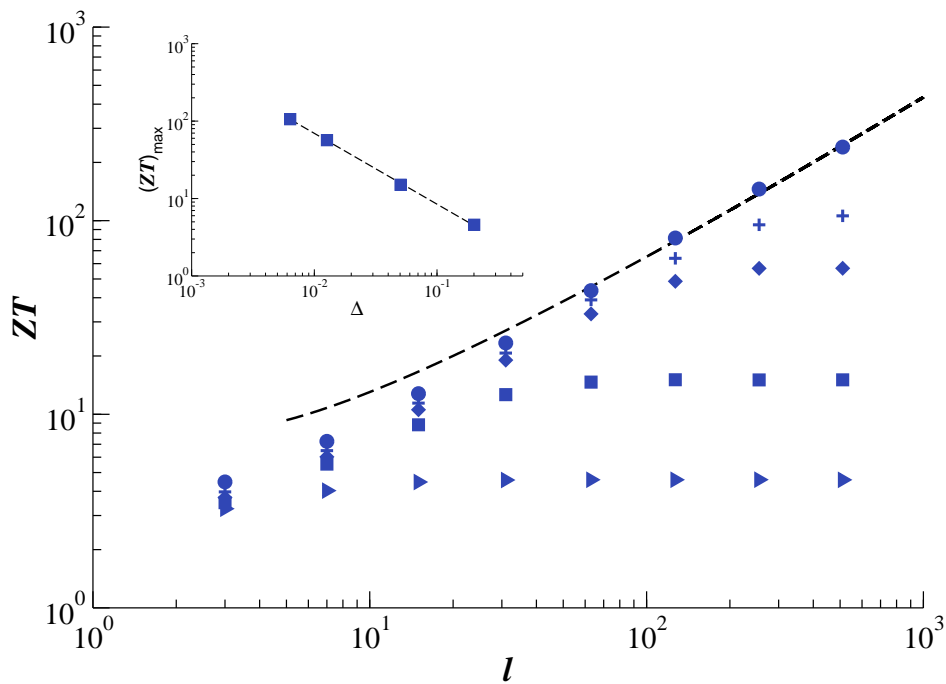}}
\caption{Thermoelectric figure-of-merit $ZT$ as a function of the
  length of the system $l$ for different thermodynamic gradients,
$\Delta=0.00625$ (circles), $0.0125$      (pluses), $0.025$
(diamonds), $0.1$      (squares) and $0.4$      (triangles)
and the other parameter values as in the caption of
Fig.~\ref{fig2}. The dashed curve stands for $\sim l/\log(l)$
In the inset, we show the maximum (saturation) value of 
$ZT$ as a function of $\Delta$. The dashed
line is a power-law fit, $(ZT)_{\rm max}=\Delta^\alpha$,
with $\alpha\approx -0.9$.
\label{fig4}}
\end{center}
\end{figure}

Through Eq.~\eref{ZT}, this  characteristic length $l^\star$ does also
determine   the  behaviour   of  the   figure  of   merit   $ZT$.   In
Fig.~\ref{fig4}  we show  $ZT$ as  a  function of  $l$, for  different
values of $\Delta$. We observe that for any value of $\Delta$, $ZT$ is
in reasonable  agreement with  an initial grow  $ l/\log(l)$ for  $l <
l^\star$  and   for  larger  sizes   saturates  to  a   maximum  value
$(ZT)_{\mathrm{max}}$.  Our results show  that as a consequence of the
existence of a single relevant  conserved quantity, the values of $ZT$
are  greatly enhanced  when the  system under  consideration  is large
enough.   Moreover, $ZT$  does  not grow  unboundedly,  but reaches  a
maximum value that grows with $\approx(1/\Delta)^{0.9}$ (see the inset
of Fig.~\ref{fig4}). The deviations at short sizes are probably due to
the  slow convergence  of the  Seebeck coefficient  to  its asymptotic
value $2$.

\begin{figure}[!h]
  \centerline{\includegraphics*[width=0.8\textwidth]{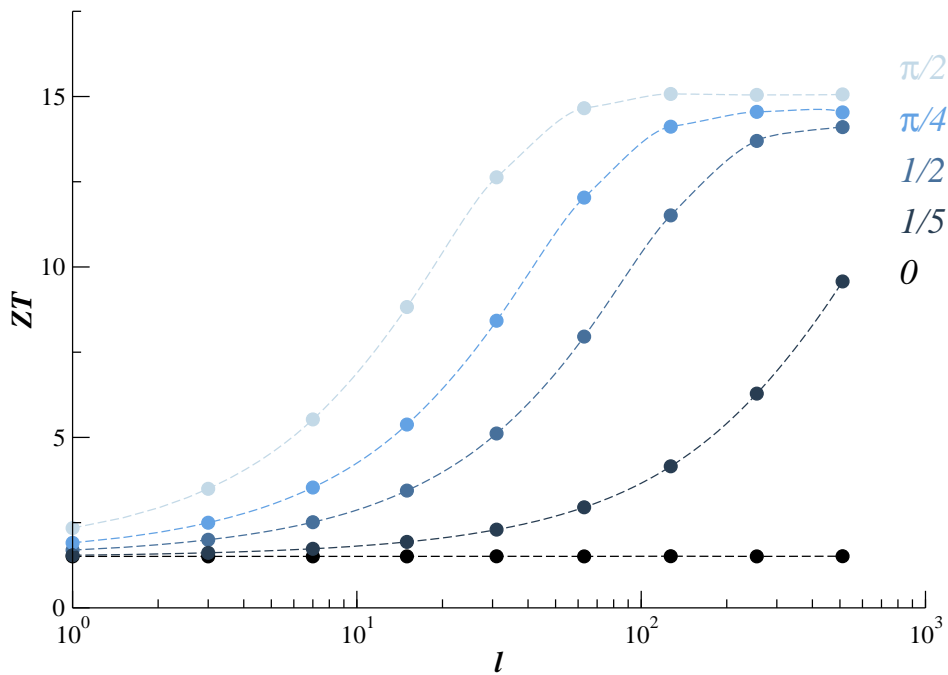}}
  \caption{Figure-of-merit  $ZT$ as a  function of  the length  of the
    system  $l$,  for  different  values of  the  collision  parameter
    $\alpha$, for  $\Delta=0.1$, and  from bottom to  top: $\alpha=0$,
    $1/5$, $1/2$, $\pi/4$ and  $\pi/2$. The other parameter values are
    as in Fig.~\ref{fig2}.}
\label{fig5}
\end{figure}

In the above discussion on the behaviour of the transport coefficients 
and the figure of merit $ZT$ as a function of $\Delta$, we should keep 
in mind that such coefficients and consequently also $ZT$ are
defined in the linear response regime, i.e. in the limit of 
small thermodynamic forces, formally for $\Delta\to 0$. 
On the other hand, we numerically computed the kinetic coefficients,
for any given $\Delta$,
via the fluxes as discussed at the beginning of Sec.~\ref{sec:discussion}.
That is to say, there is no 
saturation of $ZT$ within linear response. On the other hand, the 
numerically observed saturation (as well as the ballistic behaviour of 
$\kappa$ for $l>l^\star$) signals that the range of linear response shrinks 
with the system size when computing $\kappa$ and $ZT$.
At first sight, this failure of linear response 
for a given $\Delta$ and large $l$
appears counterintuitive,  since for  fixed $\Delta$ larger  $l$ means
smaller thermodynamic forces, and it is in the limit of small forces that 
linear response is expected to be valid.
There is actually no such problem when computing the kinetic coefficients
$L_{ij}$. As shown in Fig.~\ref{fig3} for the charge conductivity 
$\sigma=L_{\rho\rho}/T$, data at different $\Delta$ collapse on a single 
curve, showing that for all values of $\Delta$ in that figure we are 
within linear response. The problem arises when considering non-trivial 
combinations of the kinetic coefficients, as in $\kappa\propto \det ({\bm L})$
and consequently in $ZT$.
Our theory predicts the divergence of $ZT$ in the thermodynamic limit and 
$ZT$ diverges (thus leading to Carnot efficiency) if and only if the 
Onsager matrix $L$ becomes ill-conditioned, namely the condition number
$[\Tr({\bm L})]^2/\det({\bm L})$ diverges (in our model 
as $l/\log(l)$) and therefore the system 
\eref{eq:lresponse} becomes singular. That is, the charge and energy currents
become proportional, a condition commonly referred to as {\it strong coupling}, 
i.e. $J_\rho= c J_u$, the proportionality factor $c$ 
being independent of the applied thermodynamic forces.
The Carnot efficiency is obtained in such singular limit and it is in 
attaining such limit that the validity range of linear response shrinks.
Therefore, as  expected  on general  grounds, the  Carnot
efficiency is obtained only in the  limit of zero forces and 
zero currents, corresponding to
reversible transport (zero entropy production) and zero output power.

It is  worthwhile noticing that  for our model in  the non-interacting
limit the  momentum of  each particle is  conserved, meaning  that the
system is integrable and the number of conserved observables $M\propto
l$, thus diverging  in the thermodynamic limit.  As  we have discussed
at  the  end  of  section~\ref{conslaw},  one  expects  that  in  such
integrable situation,  $ZT$ does not  scale with the system  size.  To
corroborate this expectation we show in Fig.~\ref{fig5} the dependence
of  $ZT$ on  $l$ for  different  values of  the collisional  parameter
$\alpha$.   We  recall  that  at  the  collisions,  $\alpha  =  \pi/2$
corresponds  to the  most efficient  mixing of  the  particle momenta,
while $\alpha = 0$ corresponds to no interaction. As expected, for the
non-interacting  gas,  namely  for  an infinite  number  of  conserved
quantities, $ZT$  does not scale  with $l$, attaining the  value $3/2$
characteristic of  a two-dimensional ideal  gas \cite{carlos2008}. The
enhancement of $ZT$ is observed  for any value of $\alpha>0$, as
then only  the total momentum is  preserved and $M=1$.   Our data also
suggest a rather weak dependence of $(ZT)_{\mathrm{max}}$ on $\alpha$.

\subsection{Systems with noise}

The  results discussed  above show the  enhancement of
$ZT$,  and thus  of  the thermoelectric  efficiency,  in systems  with
conserved total momentum. In  real systems, however, total momentum is
never conserved due to the phonon field, the presence of impurities or
in  general  to  inelastic  scattering  events. 

In this  section we want to explore  to what extent the  break down of
total  moment conservation  modifies the  results obtained  above.  To
address this question numerically, we consider the existence of a source of
stochastic noise.  From a  physical point of  view, this  noise source
may model the interactions of the gas with the walls of the container,
or the inelastic scattering from impurities in the material.
We model the stochastic noise as follows: after a 
collision of the particles in a given cell has taken place, with probability 
$\varepsilon$ the velocities of all the particles in the cell are reflected,
namely $\vec{v}_i \rightarrow -\vec{v}_i$.
Therefore  for any  $\varepsilon>0$ the
total momentum is not longer conserved.  If $\varepsilon$ is small the
momentum conservation is weakly broken  and we want to investigate how
our results depend on the strength $\varepsilon$ of the perturbation.

\begin{figure}[!t]
  \centerline{\includegraphics*[width=1.0\textwidth]{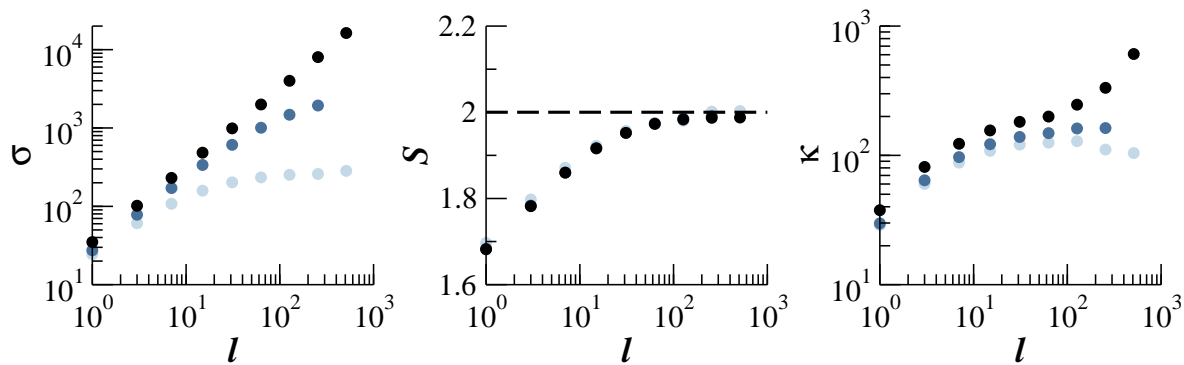}}
  \caption{The dependence of the transport coefficients on the size $l$ of
    the system, for $\Delta=0.0125$ and  different noise intensities:
     from darker to lighter, $\varepsilon =0$, $0.01$ and $0.1$.
The other parameter values are as in Fig.~\ref{fig2}.}
\label{fig6}
\end{figure}

\begin{figure}[!t]
  \centerline{\includegraphics*[width=0.8\textwidth]{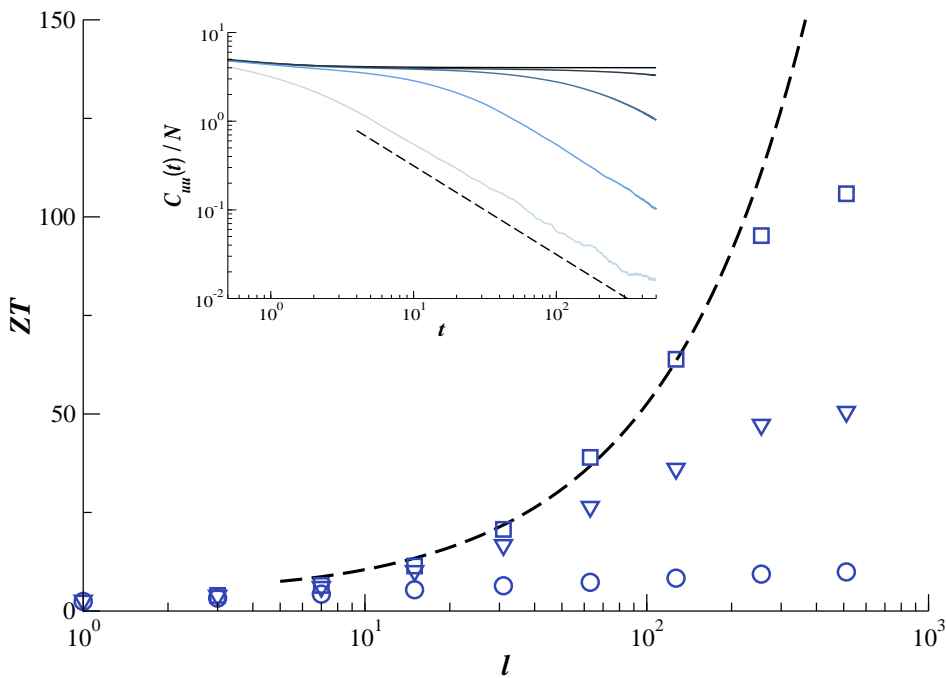}}
  \caption{Figure-of-merit  $ZT$ as a  function of  the length  of the
    system  $l$,  for   different  noise  intensities  $\varepsilon=0$
    (squares),   $0.01$   (triangles)   and  $0.1$   (circles).    The
    thermodynamic  gradient was fixed  to $\Delta=0.0125$.  The dashed
    curve   corresponds   to  the   expected linear   response  dependence   at
    $\varepsilon=0$, i.e.  $ZT\sim l/\log(l)$.  In the inset: energy
    current-current   correlation  for  different   noise  intensities
    $\varepsilon$,   for    the   same   parameter    values   as   in
    Fig.~\ref{fig6}.         From         top        to        bottom:
    $\varepsilon=0,10^{-4},10^{-3},10^{-2},10^{-1}$.  The dashed curve
    stands for $\sim 1/t$.}
\label{fig7}
\end{figure} 

In   Fig.~\ref{fig6}  we   show  the   dependence  of   the  transport
coefficients on $l$ for fixed  $\Delta$ and different strengths of the
noise $\varepsilon$. We observe that for sufficiently strong noise all
transport coefficients appear to  become independent of $l$,
as  expected in a diffusive regime in which  total momentum  is not
preserved.

More interesting is the behaviour of $ZT$ shown in Fig.~\ref{fig7}. We
see that at stronger noise,  $ZT$ becomes constant, as expected in the
diffusive regime.  From  a mathematical point of view,  the absence of
conserved quantities  ($M=0$) leads to  decaying correlation functions
and  zero  Drude coefficients  (inset  of  Fig.~\ref{fig7}). Thus  the
transport coefficients and $ZT$ become size-independent.

More  importantly,  we  see  that  when  the  convergence  toward  the
diffusive  regime is  smooth, meaning  that when  the  conservation of
total momentum  is only {\em weakly}  perturbed (small $\varepsilon$),
the enhancement of $ZT$ can  be significant.  
%The saturation value for
%$ZT$ is smaller than the corresponding $(ZT)_{\mathrm{max}}$ obtained,
%for a given $\Delta$, in the  clean case, but larger than the value of
%$ZT$ for diffusive systems.
This shows that the effect described here is robust against perturbations.

\section{Conclusions}
\label{sec:conclusions}

In summary, we have shown that in two-dimensional interacting systems,
with the interactions modeled by the multi-particle collision dynamics
method, the thermoelectric figure of merit diverges at the thermodynamic
limit. In such limit, the Carnot efficiency is obtained with zero output power. 
When noise is added to the system, $ZT$ saturates at large $l$, to values higher 
the weaker is the noise strength. 

Our findings could be relevant in situations in which the elastic mean
free path is longer than  the length scale over which interactions are
effective  in  exchanging  momenta  between the  particles.   Suitable
conditions  to  observe  the  interaction-induced enhancement  of  the
thermoelectric  figure  of  merit  might  be  found  in  high-mobility
two-dimensional electron  gases at low temperatures.   In such systems
very large elastic  mean free paths have been  reported (for instance,
up to  $28$ $\mu$m in Ref.~\cite{goldhaber}). At  low temperatures the
inelastic   mean  free   path  is   determined   by  electron-electron
interactions rather  than by phonons. It should  be therefore possible
to  find  a temperature  window  where electron-electron  interactions
dominate, i.e. are effective on  a scale smaller than the elastic mean
free path and  are dominant over phonon effects. It  would be, however,
highly desirable  to test  our arguments in  such regime, by  means of
numerical simulations of quantum systems.

\ack
We acknowledge support by MIUR-PRIN and by Regione Lombardia. CMM is partially supported by the European Research Council, the Academy of Finland, and by the MICINN (Spain) grant  MTM2012-39101-C02-01.

\end{document}